\documentclass[superscriptaddress,twocolumn,secnumarabic,
amssymb,amsmath,nobibnotes,aps,prd,showpacs,nofootinbib]{revtex4}%
\usepackage{graphicx}
\usepackage{epsf}
\usepackage{bm}
\usepackage{amsmath}
\usepackage{amsfonts}
\usepackage{amssymb}%
\providecommand{\U}[1]{\protect\rule{.1in}{.1in}}
\newcommand{\be}{\begin{equation}}
\newcommand{\ee}{\end{equation}}
\newcommand{\CC}{\Lambda}

\newcommand{\Omo}{\Omega_{m0}}

\newcommand{\OLo}{\Omega_{\Lambda 0}}

\newcommand{\wm}{\omega_m}
\newcommand{\rco}{\rho_{c0}}
\newcommand{\rmo}{\rho_{m0}}
\newcommand{\xim}{\xi_{\omega}}
\newcommand{\rmm}{\rho_{m}}
\newcommand{\PLB}[3]{{ Phys. Lett. } {\bf B#1} (#2)  {#3}}

\newcommand{\CQG}[3]{{ Class. Quant. Grav. } {\bf #1} (#2) {#3}}
\newcommand{\JHEP}[3]{ {JHEP} {#1} (#2)  {#3}}
\newcommand{\PR}[3]{{ Phys. Rep. } {\bf #1} (#2)  {#3}}
\newcommand{\CH}{C_H}
\newcommand{\rplu}{r_{+}}
\newcommand{\rmin}{r_{-}}
\newcommand{\CHd}{C_{\dot{H}}}
\newcommand{\Hd}{\dot{H}}
\newcommand{\pmr}{p_m}
\newcommand{\tetm}{\theta_{m}}
\newcommand{\rLo}{\rho_{\CC 0}}
\newcommand{\rmr}{\rho_m}

\newcommand{\rDE}{\rho_{\rm DE}}
\newcommand{\mincir}{\raise
-3.truept\hbox{\rlap{\hbox{$\sim$}}\raise4.truept\hbox{$<$}\ }}
\newcommand{\magcir}{\raise
-3.truept\hbox{\rlap{\hbox{$\sim$}}\raise4.truept\hbox{$>$}\ }}
\newcommand{\newtext}[1]{\text{#1}}

\newcommand{\rL}{\rho_{\Lambda}}

\newcommand{\wDE}{\omega_{\rm DE}}

\begin{document}
\title{Entropic-force dark energy reconsidered}

\author{Spyros Basilakos}
\affiliation{Academy of Athens, Research Center for Astronomy and
Applied Mathematics, Soranou Efesiou 4, 11527, Athens, Greece}

\author{Joan Sol\`{a}}
\affiliation{High Energy Physics Group, Dept. ECM, Univ. de
Barcelona, Av. Diagonal 647, E-08028 Barcelona, Catalonia, Spain}
\affiliation{and Institut de Ci{\`e}ncies del Cosmos (ICC),  Univ.
de Barcelona}

\begin{abstract}
We reconsider the entropic-force model in which both kind of Hubble
terms ${\dot H}$ and $H^{2}$ appear in the effective dark energy
(DE) density affecting the evolution of the main cosmological
functions, namely the scale factor, deceleration parameter, matter
density and growth of linear matter perturbations. However, we find
that the entropic-force model is not viable at the background and
perturbation levels due to the fact that the entropic formulation
does not add a constant term in the Friedmann equations.
On the other hand, if on mere phenomenological grounds we replace
the ${\dot H}$ dependence of the effective DE density with a linear
term $H$ without including a constant additive term, we find that
the transition from deceleration to acceleration becomes possible
but the recent structure formation data \newtext{strongly disfavors}
this cosmological scenario. Finally, we briefly compare the
entropic-force models with some related DE models (based on
dynamical vacuum energy) which overcome these difficulties and are
compatible with the present observations.

\end{abstract}
\date{\today}

 \maketitle

\hyphenation{tho-rou-ghly in-te-gra-ting e-vol-ving con-si-de-ring
ta-king me-tho-do-lo-gy fi-gu-re}

\section{Introduction}
The discovery of the cosmic acceleration (see
\cite{Spergel07,essence,Kowal08,Hic09,komatsu08,LJC09,BasPli10,Ade13}
and references therein) has opened a new window in trying to
understand the universe. Despite the mounting observational evidence
on the existence of the accelerated expansion of the universe, its
nature and fundamental origin is still an open question challenging
the very foundations of theoretical physics. While the simplest
possibility is a rigid cosmological constant (CC) term $\CC$ for the
entire history of the universe, with the advent of quantum theory
and quantum field theory (QFT) the $\CC$-term has been associated
with the vacuum energy density: $\rL = \CC/(8\pi G)$, and this
association is at the root of the acute cosmological constant
problem\,\cite{CCproblem}, perhaps one of the most pressing
conundrums of fundamental physics ever -- see e.g.
\cite{JSP-CCReview13} for a recent review.

The difficulties inherent to the CC problem suggest to adopt a more
dynamical perspective. This has led to various forms of dynamical
dark energy (DE) as a possible substitute for the rigid CC term.
Along these lines one may consider a general mechanism that is
responsible for cosmic acceleration which is based either on a
modified theory of gravity or on the existence of some sort of DE
which is related with the existence of new fields in nature.

Perhaps the most popular idea in the past was to find a scalar field
capable of dynamically adjusting the vacuum energy to zero (or, in
fact, to the tiny number $\rLo\sim 10^{-47}$ GeV$^4$ that has been
measured in recent times), with the hope to solve the old CC
problem\,\cite{CCproblem}. The first attempts in this direction are
quite old\,\cite{Dolgov82,OldDynamAdjust,PSW}. It is only in more
recent times that this approach took the current popular form of
quintessence\,\cite{Quintessence,QuintessenceReview} and was mainly
applied to explain the possible dynamical character of the DE, with
an eye at solving the ``cosmic coincidence'' problem. The DE ideas
since then have branched off into multiple different formulations,
see e.g. \cite{curvature,mauro,report,repsergei,
Oze87,Lambdat,Brax:1999gp,KAM,fein02,Caldwell,Bento03,chime04,Linder2004,LSS08,
Brookfield:2005td,Boehmer:2007qa,Starobinsky-2007,Ame10} and
references therein. In particular, we have also the dynamical models
of the vacuum energy based on the renormalization group (RG)
approach \,\cite{RGCosmology1,RGCosmology2,Fossil07,Bas09c,GSBP11}.
As we will see, these models bare a close (but distinctive) relation
with the entropic-force models.

Amongst the variety of DE models, the so-called entropic-force dark
energy has recently gained a lot of attention in cosmology. The
notion of entropic-force was suggested by
Verlinde\,\cite{Verlinde10}, who proposed that the gravitational
field equations can be derived from the second law of thermodynamics
in a way that would render the gravity force quite literally as a
kind of ``entropic force'', i.e. a force related to the change of
entropy. Its implementation in cosmology\,\cite{Easson10,Easson10b}
is based on the assumption that the horizon can play the role of
holographic screen\,\footnote{Let us mention that in
Jacobson's\,\cite{Jacobson95} and Padmanabhan's approaches to
entropic gravity\,\cite{Padmanabhan10}, the formulation is much more
general and is not affected at all by our analysis here, which is
restricted to the specific ``entropic-force''
version\,\cite{Verlinde10} and its cosmological implementation
in\,\cite{Easson10,Easson10b}. See also Ref. \cite{Visser11} for the
difficulties of interpreting entropic forces in Newtonian gravity.}
(see also \cite{Mia011}).

Such screen would induce a force $\textbf{F}=T\,\nabla S$ on a
nearby test particle, where $T$ is the temperature of the screen and
$\nabla S$ is the change of entropy associated with the information
contained in it (which involves a large number of d.o.f.). The
change of entropy when the radius of the Hubble horizon,  $R_H=1/H$,
increases by $dr$ is simply $dS_H=2\pi\,(R_H\,k_B\,M_P^2)\,dr$,
where $M_P$ is the Planck mass and $k_B$ is Boltmann's constant.
From the pressure exerted by the entropic force on the cosmological
expansion, $P=F/A=-(T/A)\,dS_H/dr$, and estimating the horizon
temperature by the de Sitter temperature, $T=(\hbar/k_B)(H/2\pi)$,
one finally obtains $P=-\,M_P^2 H^2/4\pi$. The minus sign in the
pressure is of course the characteristic feature of the accelerated
expansion in this entropic version.  This framework therefore
suggests that the entropic force leads to an effective DE density
$\rDE$ which is dynamical and evolves as the square of the Hubble
rate: $\rDE\propto H^2$.

In this formulation the DE does not exist as an exotic energy
component of the universe, but as an effective force acting outwards
the cosmic horizon, thereby accelerating the evolution of the
universe. This particular scenario is the so called ``entropic-force
cosmology'' and was first proposed in
Ref.\,\cite{Easson10,Easson10b} and later on discussed by various
authors -- see e.g. \,\cite{Casadio10,Entropic-others,Visser11,
Polarski,Japanese12,Japanese3}.

One crucial question is what classes of entropic-force models are
allowed in cosmological studies. Specifically, we are interested in
testing the validity of the entropic-force DE both at the background
and cosmic perturbations levels. In order to do so we need to define
the Hubble parameter and the growth of matter perturbations in the
linear regime as a function of redshift and then to check whether
the above functions could lead to reliable cosmological results.

The layout of the article is the following. In Sec. 2, we present
the main ingredients of the dynamical problem under study. In Secs.
3 and 4 we provide the analytical solutions for the Hubble and
growth factor respectively. Finally, we draw our conclusions in Sec.
5.

\section{Entropic-force models and effective dark energy}
\label{sect:entropicforce} The nature of the entropic-force
models\,\cite{Verlinde10} is
essentially connected with a surface effect from the horizon. From
the latter one may utilize the gravitational action for space-times
with boundaries\,\cite{Hawking96}. This is achieved by adding the
boundary action term $S_B$ to the standard Einstein-Hilbert action,
$S_{EH}$, namely:
\begin{equation}\label{IB}
S_{EH}+S_B=\frac{1}{16\pi G}\int_{{\cal M}}\,d^4
x\sqrt{|g|}\,R\,+\,\frac{1}{8\pi G}\int_{\partial{\cal M}}\,d^3
y\sqrt{|h|}\,K\,.
\end{equation}
Here $R$ is the Ricci scalar, $h$ is the determinant of the metric
$h_{ab}$ on the boundary $\partial{\cal M}$, induced by the bulk
metric $g_{\mu\nu}$ of ${\cal M}$, and $y^a$ are the coordinates on
$\partial{\cal M}$. Furthermore, $K$ is the trace of the second
fundamental form (or extrinsic curvature); if $n^{\mu}$ is the
normal on the boundary, it can be written as
$K=\nabla_{\mu}n^{\mu}$. The overall action is $S=S_{EH}+S_{B}+S_m$,
where $S_m$ represents the ordinary matter contribution. From the
technical point of view, the precise definition of the boundary term
$S_B$ should actually include an overall sign, which is plus or
minus depending on whether the hyper-surface $\partial{\cal M}$ is
space-like ($n^{\mu}n_{\mu}=+1$) or time-like ($n^{\mu}n_{\mu}=-1$),
respectively. We exclude null surfaces for this consideration.

The precise coefficient in front of the boundary integral  $S_{B}$
is chosen in such a way that the surface terms generated from the
metric variation of $S_{EH}$ are exactly canceled by the metric
variation of $S_{B}$, provided the variation $\delta g^{\mu\nu}$ is
performed in such a way that it vanishes on $\partial{\cal M}$, i.e.
provided the induced metric $h_{ab}$ on the boundary is held fixed.
It follows that, in the presence of $S_B$, the standard form of
Einstein's equations is preserved even if the space-time has
boundaries.

As the surface terms emerging from the variation of $S_{EH}$ are
canceled by $\delta S_B$, one may assume that if the total action
would not contain $S_B$ the contribution of the aforementioned
surface terms to the field equations would be of the order of the
effect induced by $S_B$,  estimated as $R$ times the
 prefactor $1/(8\pi G)$ in $S_B$. In the FLRW metric with flat space slices this reads $(12 H^2+6\dot{H})/(8\pi G)$,
with $H=\dot{a}/a$ and $\dot{H}=dH/dt$. This is presumably the kind
of consideration made by Easson et al. \cite{Easson10}. They also
argued that since this is probably just a rough estimate of the
effect, one should generalize the corresponding acceleration
equation for the scale factor in the form:
\begin{equation}\label{entropica}
\frac{\ddot{a}}{a}=-\frac{4\pi\,G}{3}\,(1+3\wm)\,\rmr+\CHd\,\Hd\,+\CH\,H^2
\end{equation}
where $\wm=\pmr/\rmr$ is the equation of state (EoS) for the matter
fluid (with $\wm=0$ and $1/3$ for non-relativistic and relativistic
matter, respectively). The second cosmological equation is the
generalized Friedmann's equation:
\begin{equation}\label{GeneralizedFriedmann}
H^2\equiv\left(\frac{\dot{a}}{a}\right)=\frac{8\pi\,G}{3}\,[\rmr(t)+\rDE(t)]\,,
\end{equation}
where we always assume flat space metric. In the above expression,
we have defined
\begin{equation}\label{EffectiveDE}
\rDE(t)=\frac{3}{8\pi G}\left[\CHd\,\Hd(t)+\CH H^2(t)\,\right]\,.
\end{equation}
It plays the role of effective DE for the generalized cosmological
model. Despite the two equations above imply $\wDE=P_{\rm
DE}/\rDE\equiv -1$, the entropic-force model is not  a time varying
vacuum model since it does not have a $\CC$CDM limit, that is to
say, we cannot obtain a constant $\rDE$ behavior at any time in its
Hubble expansion history.

Let us emphasize at this point an issue previously mentioned in
passing, but that is important when we consider the status of the
basic equations (\ref{entropica}) and (\ref{EffectiveDE}). These
equations are not necessary derived from first principles, such as a
fundamental action. As a matter of fact, in the most general
entropic-holographic formulations\,\cite{Jacobson95,Padmanabhan10}
one considers that the gravitational field equations are not
necessarily inferred from a fundamental action at the present
macroscopic level of description. The field equations can
nevertheless provide, in principle, a fully satisfactory account of
all the basic phenomena known to date. In such general ``emergent
gravity formulation'', the ultimate origin of gravity is claimed to
lie in some fundamental degrees of freedom quite different from the
metric variables, namely degrees of freedom which are completely
unknown to us at present. In this sense, the field equations under
discussion are also thought of as just effective field equations
without having a known fundamental action behind. However, we should
also remark that these particular equations pertain to a restricted
formulation of the general entropic approach called
``entropic-force'' gravity, and it is our aim to discuss the
phenomenological status of such restricted version of the entropic
approach concerning the cosmological implications first suggested in
Ref.\cite{Easson10}. In this sense our conclusions concerning the
phenomenological viability of these models cannot be extended beyond
the domain of this formulation.

Implicit herein is the fact that matter is \emph{not} conserved in
the entropic-force models. Indeed, Eqs.(\ref{entropica}) and
(\ref{GeneralizedFriedmann}) can be used to show that matter and the
effective DE (\ref{EffectiveDE}) exchange energy through the
generalized conservation law:
\begin{equation}\label{Bronstein}
\dot{\rho}_{\rm DE}+\dot{\rho}_m+3(1+\wm)\,H\,\rmr=0\,.
\end{equation}
This equation is therefore not an independent one, but it is useful
to show explicitly that there is a continuous exchange of energy
between matter and the generalized DE given by
Eq.(\ref{EffectiveDE}). Such equation is actually enforced by the
Bianchi identity, which implies $\nabla^{\mu}\tilde{T}_{\mu\nu}=0$
for the total energy-momentum tensor $\tilde{T}_{\mu\nu}$, obtained
from the ordinary matter contribution, ${T}_{\mu\nu}$, plus the
variable DE density (\ref{EffectiveDE}).
This is in contradistinction to what it was assumed in
\cite{Easson10}. Let us note that in \cite{Easson10b} the treatment
of the covariant matter conservation law was different from the
previous one by the same authors in \cite{Easson10}, and the
modified law was presented in the form
\begin{equation}\label{correctedlaw}
\dot{\rho}_m+3\,(1-C_H)(1+\wm)\,H\,\rmr=0\,.
\end{equation}
However, this equation is only approximate. It can be easily derived
e.g. from our equations (\ref{GeneralizedFriedmann}) and
(\ref{EffectiveDE}) after neglecting the $\CHd$ contribution.
Below we will present the exact solution for $\rmr$ consistent with
the full conservation equation (\ref{Bronstein}), and then
(\ref{correctedlaw}) will ensue only as a particular case.

It is not difficult to show that equations
(\ref{entropica})-(\ref{EffectiveDE})
can be combined to produce the following differential equation for the
Hubble rate:
\begin{equation}\label{BasicdiffEquation}
\dot{H}+\frac32\,\left(1+\wm\right)\,\frac{1-C_{H}}{1-\alpha}\,H^2
=0\,,
\end{equation}
where we have defined $\alpha=\frac32\,\left(1+\wm\right)\,\CHd$.
Using the transformation from the cosmic time $t$ to the scale
factor $a$ through $d/dt=aH(a)d/da$ in
Eq.\,(\ref{BasicdiffEquation}) and defining $E(a)=H(a)/H_{0}$ we can
easily present the solution of Eq.(\ref{BasicdiffEquation}) as
follows:
\begin{equation}\label{Hentropic}
E(z)=(1+z)^{3\xim/2}\,,
\end{equation}
where $z=a(z)^{-1}-1$ is the redshift, and we have defined the
important parameters
\begin{eqnarray}\label{xiparameter}
 \xim&=&\xi(1+\wm)\nonumber\\
\xi&=&\frac{1-C_{H}}{1-\alpha}\,.
\end{eqnarray}
If we consider $C_{H}=3\CHd/2$ and $\wm=0$, then $\xim=\xi=1$ and
the above solution boils down to the Einstein de Sitter model.

The matter density in the entropic-force model is evaluated  from
\begin{eqnarray}\label{eq:matterdens}
\rho_m(z)&=&-\frac{1}{8\pi G
(1+\wm)}\frac{dH^2(a)}{d{\rm ln}a}\nonumber\\
&=&\frac{1+z}{3(1+\wm)}\,\rco\,\frac{dE^2(z)}{dz}\,.
\end{eqnarray}
Computing this expression and subsequently using it for solving
Eq.\, (\ref{Bronstein}) we can obtain the DE density as well. The
final results are:
\begin{equation}\label{rhomrhoLentropic}
\rmr(z)=\rco\,\frac{1-C_{H}}{1-\alpha}\,(1+z)^{3\xim}\equiv\rmo\,(1+z)^{3\xim}
\end{equation}
\begin{equation}\label{rhomrhoLentropic1}
\rDE(z)=\rco\,\frac{C_{H}-\alpha}{1-\alpha}\,(1+z)^{3\xim} \equiv
\rho_{{\rm DE},0}\,(1+z)^{3\xim}\,.
\end{equation}

In these expressions we identify in the matter era ($\wm=0$,
$\xim=\xi$)
\begin{equation}\label{OmoOLoentropic1}
\Omo=\frac{\rmo}{\rco}=\frac{1-C_{H}}{1-\alpha}=\xi\,,
\end{equation}
\begin{equation}\label{OmoOLoentropic2}
\Omega_{{\rm DE},0}=\frac{\rho_{{\rm
DE},0}}{\rco}=\frac{C_{H}-\alpha}{1-\alpha}=1-\xi\,,
\end{equation}
which clearly satisfy $\Omo+\Omega_{{\rm DE},0}=1$, as expected.
Furthermore, if one neglects $\CHd$ (hence $\alpha\simeq 0$) we have
$3\xim\simeq3(1-C_H)(1+\wm)$, and then we recover -- as a particular
case-- the solution of the approximate Eq.\,(\ref{correctedlaw})
pointed out by the authors of Ref.\,\cite{Easson10b}.
Finally, using Eq.(\ref{Hentropic}) the deceleration parameter
$q=-\ddot{a}/a H^2$ can be equivalently computed from
\begin{equation}\label{eq:qnu}
q=-1-\frac{d{\rm ln}H}{d{\rm
ln}a}=-1+\frac{1+z}{E(z)}\frac{dE(z)}{dz}= -1+\frac{3\xim}{2} \;.
\end{equation}
It is evident that this cosmological model has no inflection point
in its cosmic history, i.e. a point where deceleration can change
into acceleration. This is because Eq.(\ref{eq:qnu}) is redshift
independent and hence the deceleration parameter maintains sign
throughout the cosmic history. As a result the universe always
accelerates or always decelerates depending on the value of $\xim$.
In the matter dominated era, namely $\wm=0$ and $\xim=\xi$ , the
accelerated expansion of the universe ($q<0$) is obtained for
$\xi<\frac{2}{3}$. But in such case we would be also admitting that
the universe has been accelerating forever, which is of course
difficult to accept. Clearly, this feature is  the main drawback of
this model.

Finally, we would like to mention once more that in this section we
have discussed the background evolution via Eqs.(\ref{entropica}),
(\ref{GeneralizedFriedmann}) of the DE models considered by the
author of the entropic force models \cite{Easson10}. Although one
could define the DE in different ways, our aim was to show that
using the nominal approach employed by the original authors the
entropic-force model in which both kind of Hubble terms ${\dot H}$
and $H^{2}$ appear in the effective dark energy (DE) is in trouble.
In this context, the DE pressure is assumed to satisfy the vacuum
equation of state, namely $P_{\rm DE}=-\rDE$. This is important
since in the original model of \cite{Easson10} the above equation of
state was not mentioned. In our case we make this assumption, which
is the minimum one we can do. In fact one can show that if another
equation of state would be used the kind of problems that the
entropic model has would remain. For example, if the EoS parameter
for the DE would take some arbitrary constant value $\omega_{DE}$
different from  minus one, this would just change the coefficient of
$H^2$ in Eq.\,(\ref{BasicdiffEquation}), but the solution
(\ref{Hentropic}) would take the same form (with a slightly
different expression for $\xi_\omega$) and hence the deceleration
parameter in Eq.\,(\ref{eq:qnu}) would still be independent of the
redshift. As a result the absence of the inflexion point between
deceleration and acceleration would persist.

\section{Linear matter perturbations}
\label{sect:Generalperturbations} Although finding the background
cosmological solution is of course important for our study, no less
important is to investigate the structure formation properties, as
they  play an essential role in the cosmic history. If the
entropic-force model discussed in the previous section would e.g. be
the late-time effective behavior of a more complete and fundamental
model, we could still ask ourselves quite licitly if the late-time
behavior of that model is compatible with the structure formation
data (which is collected precisely in that relevant period).

In this section we provide  a detailed answer to such question. To
this end we discuss the perturbations in the presence of an
entropic-force DE. It will suffice to consider perturbations only
for matter, but we have to incorporate the dynamical character of
the effective DE in the matter perturbation equations. In other
words, the entropic-force DE is considered variable, but homogeneous
in first approximation.

The general study of the linear perturbation equations for a
multicomponent fluid has been addressed e.g. in \cite{Grande08} and
we will use the formulation in that paper here -- see equations
(17), (25) and (27) of that reference (see also \cite{Bas09c},
\cite{Arc94}, \cite{Borg}).
Let us apply those perturbation equations for a system composed of
the effective (entropic-force) DE fluid and the matter fluid
$\rmr=\rmr(t)$, assuming that there are matter perturbations
$\delta\rmr$ but no perturbation in $\rDE$. Let us provisionally
note here that the inclusion of perturbations of the DE component in
this kind of scenarios, with and without matter conservation, is
also possible\,\cite{GSBP11,PerturbVacuum1,PerturbVacuum2}, but it
is not necessary for the present study. We shall further comment
below on the viability of this approach.

In this section (and for the entire discussion on perturbations) we
just set $\wm=0$, corresponding to non-relativistic matter in the
epoch of structure formation.  The relevant system of first order
differential equations reads:
\begin{eqnarray}\label{diffsystem}
&&\dot{\hat{h}}+2\,H\hat{h}= 8\pi G\delta\rmr\nonumber\\
&& \delta\dot{\rho}_m+\rmr \left(\tetm-\frac{\hat{h}}{2}\right)+3\,H\,\delta\rmr= 0 \\
&&\rmr\dot{\theta}_m+\left(\dot{\rho}_m+5H\rmr\right)\,\tetm =0\,,
\nonumber
\end{eqnarray}
where $\hat{h}$ is minus the time derivative of the trace of the
metric perturbation $\delta g_{\mu\nu}$, and $\tetm$ the divergence
of the perturbed matter velocity\,\cite{Grande08}.  The last
equation of the system (\ref{diffsystem}) can be readily rewritten
as
\begin{equation}\label{difftheta}
\dot{\theta}_m+\left(2H+Q\right)\tetm=0\,,
\end{equation}
where we have used Eq.\, (\ref{Bronstein}) for $\wm=0$, and defined
\begin{equation}\label{defQ}
Q\equiv\frac{\dot{\rho}_m+3\,H\,\rmr}{\rmr}=-\frac{\dot{\rho}_{\rm
DE}}{\rmr}\,.
\end{equation}
Introducing the density contrast $D\equiv\delta\rmr/\rmr$, it is not
difficult to show that the first two equations of the system
(\ref{diffsystem}) combined with (\ref{difftheta}) provide the
following second order differential equation for $D$:
\begin{eqnarray}\label{diffeqD}
\ddot{D}+\left(2H+Q\right)\,\dot{D}-\left(4\pi
G\rmr-2HQ-\dot{Q}\right)\,D\nonumber\\
=Q\,\tetm\,.
\end{eqnarray}
This equation constitutes a generalization of the basic equation for
the matter perturbations in the presence of a dynamical DE density.
In the particular case when this term is strictly constant we have
$Q=0$ and the above equation shrinks to the standard
one\,\cite{PeeblesDodelson}
\begin{equation}\label{diffeqDCC}
\ddot{D}+2H\,\dot{D}-4\pi G\rmr\,D=0,
\end{equation}
which is valid both for CDM and $\CC$CDM cosmologies with the
corresponding  Hubble function. However, when the DE is time
evolving, even if it is perfectly smooth, the correct equation is no
longer (\ref{diffeqDCC}) but (\ref{diffeqD}), a fact which is
somehow missed in some approaches in the literature.

The homogeneous version of (\ref{diffeqD}) (i.e. with its
\textit{r.h.s.} equated to zero) was first obtained using a
Newtonian approach\,\cite{Arc94}. Herein we have generalized it
within the relativistic formulation in order to discuss if the
homogeneous version can also be applied for the models under
consideration.  First of all, we note that if $|Q|< 2H$ for all $t$,
then
\begin{equation}\label{integral}
\varphi(t)\equiv\int (2H+Q)dt>0
\end{equation}
in any integration interval. Thus, if $\varphi(t)$ increases with
time the solution of (\ref{difftheta}) is a decaying one:
\begin{equation}\label{decaying}
\tetm(t)\propto e^{-\varphi(t)}\to 0\,.
\end{equation}
\phantom{x}

\noindent In all these cases the product function $Q\,\tetm$ on the
\textit{r.h.s.} of (\ref{diffeqD}) can be safely neglected.  For the
class of DE models of the form (\ref{EffectiveDE}) we can estimate,
from the definition (\ref{defQ}) of $Q$, that
\begin{equation}\label{QoverH}
\frac{|Q|}{H}\sim\frac{|C_{\dot{H}}\,\ddot{H}+C_{H}\,H\,\dot{H}|}{H^3}=\left|{\cal
O}(C_{\dot H})+{\cal O}(C_{H})\right|\,.
\end{equation}
To substantiate the last step, let us differentiate the identity
$\dot{H}+H^2=-q\,H^2$. Using $q=-\ddot{a}/aH^{2}$
and taking into account that $q\simeq const.$ in each epoch, it is
easy to see that $\ddot{H}\simeq 2(q+1)^2\,H^3$. Together with
$H\,\dot{H}=-(q+1)\,H^3$, these relations provide immediately the
estimate (\ref{QoverH}). It basically tells us that ${|Q|}/{H}\ll1$.
Of course this relation holds good insofar as $C_H$ and $\CHd$ are
expected to be small dimensionless coefficients in the
entropic-force scenario.
Therefore, Eq.\,(\ref{decaying}) is warranted and we conclude that
for the entire class of DE cosmologies (\ref{EffectiveDE}) we can
virtually ignore the \textit{r.h.s.} of the differential Eq.
(\ref{diffeqD}). Hereafter we set it to zero.

For convenience we rewrite the homogeneous form of (\ref{diffeqD})
in terms of the scale factor as independent variable, which we will
use shortly. After straightforward algebra we arrive at the
following expression:
\begin{widetext}
\begin{eqnarray}
\label{diffeqDa}
D_{aa}(a)+\left[\frac{3}{a}+\frac{H_{a}(a)}{H(a)}+\frac{Q(a)}{aH(a)}\right]\,D_{a}(a)-\left[\frac{4\pi
G\rmr(a)}{H^2(a)}-\frac{2Q(a)}{H(a)}-a\frac{Q_{a}(a)}{H(a)}\right]\,\frac{D(a)}{a^2}=0\,,
\end{eqnarray}
\end{widetext}
where $X_{a}=dX/da$ and $X_{aa}=d^{2}X/da^{2}$. In particular,
notice that for $C_H=\CHd=0$ Eq.\,(\ref{defQ}) tell us that $Q=0$
and then (\ref{diffeqDa}) reduces to the standard perturbation
equation \cite{PeeblesDodelson}.

Let us now comment on the effective approach to perturbations based
on Eq.\,(\ref{diffeqDa}). Use of this equation  implies a treatment
of matter perturbations in which the dynamical vacuum energy density
is included only at the background level, that is to say, we do not consider the vacuum energy perturbations. We proceed within this
approximation because it is the simplest possible way to discuss the
perturbations of the entropic models under consideration, which are
themselves effective models of the dark energy. We keep in mind the results from previous calculations of cosmic perturbations in models where the dark energy component was also parameterized as a power series of the Hubble function, with or without exchange of energy with the matter
component (and including in some cases a possible time variation of
the gravitational coupling). In these studies, see
Ref.\,\cite{PerturbVacuum1,PerturbVacuum2}, it is found that the
full perturbative treatment of matter and dark energy perturbations
in two different gauges (synchronous and conformal Newtonian) leads to
consistent results. Furthermore, the confrontation of the models with the
basic cosmological data produces similar results to those obtained from the effective treatment where the dynamical character of the dark energy
is encoded as in Eq.\,(\ref{diffeqDa})\,\cite{Bas09c,GSBP11}. We do
not exclude, however, that ambiguities can still be present in the
perturbation equations and the assumptions made in their derivation.
As indicated in the aforementioned references, the gauge issues can
be important, and in our case the lack of a manifest covariant
formulation can play also a role in this issue.  For the usual scales
explored in the analysis of the matter power spectrum, at sub-Hubble
domains, we assume that our effective approach reflects the basic
features.


Using (\ref{Hentropic}), (\ref{rhomrhoLentropic}) and
(\ref{rhomrhoLentropic1}), Eq.\,(\ref{diffeqDa}) can be cast as
follows:
\begin{equation}\label{eq:perturbEntropic}
a^{2}D_{aa}(a)+\frac32\,A a D_{a}(a)-\frac32\,B D(a)=0\,,
\end{equation}
where
\begin{eqnarray}\label{eq:defAB}
A&=&\frac{1+3C_{H}-4\alpha}{1-\alpha}=4-3\xi\,,\nonumber\\
B&=&\frac{(1+C_{H})(1-3C_{H})+4\alpha(2C_{H}-\alpha)}{(1-\alpha)^2}\nonumber\\
&=&(3\xi-2)\,(2-\xi)\,.
\end{eqnarray}
We recall that $\xim=\xi$ ($\wm=0$) in the structure formation
epoch, with $\xi$ defined in (\ref{xiparameter}).

As a particular case we consider the situation corresponding to
$\CHd=0$ ($\alpha=0$). In this case $\xi=1-C_{H}$ and we immediately
recover the results obtained for the $\rDE\sim H^2$ model studied
in\,\cite{Bas09c}. The corresponding  perturbations
Eq.(\ref{eq:perturbEntropic}) takes the  form
\begin{eqnarray}\label{eq:perturbH2}
&&a^2{D}_{aa}(a)+\frac32\,(1+3\CH)\,a\,D_a(a)\\
&&\phantom{xxxxx}-\frac32\,(1+\CH)(1-3\CH)\,a^2\,D(a)=0\,.\nonumber
\end{eqnarray}
The general solution reads
\begin{equation}\label{eq:solperturbH2}
D(a)=C_1\ a^{1-3\CH}+C_2\,a^{-3(1+\CH)/2}\,.
\end{equation}
As we can see, for $|\CH|<1$ (the expected situation) only the
$a^{1-3\CH}$ mode is a growing one, provided $\CH<1/3$ (equivalent
to $\xi>2/3$ for $\CHd=0$). However, this option for a growing mode
is incompatible with having positive acceleration (or $q<0$) for
this cosmology, as we have shown in  the previous section -- see
Eq.\,\eqref{eq:qnu}.

As a matter of fact, this situation will replicate for the entire
class of entropic models of the form Eq.(\ref{EffectiveDE}) and not
only for those with $\CHd=0$. Indeed, the general solution of
(\ref{eq:perturbEntropic}) is a linear combination of the modes
$a^{\rplu}$ and $a^{\rmin}$, with
\begin{equation}\label{eq:rps}
r_{\pm}=\frac12\left(1-\frac32 A\pm\sqrt{\Delta}\ \right)\,,\ \ \
\Delta\equiv \left(1-\frac32 A\right)^2+6B\,.
\end{equation}
We can check that for $\CHd=0$ these modes reduce to the ones found
in the particular case (\ref{eq:solperturbH2}).

In order to avoid oscillatory solutions in the general case
$\CHd\neq 0$, we will consider only situations in which the
discriminant $\Delta>0$. A growing mode solution of
(\ref{eq:perturbEntropic}) will exist if $\rplu>0$ and/or $\rmin>0$.
We have two possibilities: i) if $B>0$, then $\rplu>0$ and $\rmin<0$
irrespective of the sign of $1-3A/2$, so we have one growing mode;
and ii) if $B<0$ and $1-3A/2>0$, we have two growing modes:
$\rplu>0$ and $\rmin>0$. For completeness, we note that if $B<0$ and
$1-3A/2<0$, then $\rplu<0$ and $\rmin<0$, and in this case we have
no growing modes at all.

Let us now analyze if any of the two possibilities i) or ii) can be
realized in practice. From the explicit form of  $B$ as a function
of $\xi$ in Eq.\,(\ref{eq:defAB}), we can see immediately that $B>0$
is equivalent to require $2/3<\xi<2$. However, this is incompatible
with the acceleration condition $\xi<2/3$. Thus, $B>0$ is not
acceptable.

Similarly, let us consider if the scenario ii) is feasible. The
condition $B<0$ implies either $\xi<2/3$ or $\xi>2$. Only the first
case is compatible with the positive acceleration condition
$\xi<2/3$. However, let us recall that scenario ii) requires the
additional inequality $1-3A/2>0$ in order to insure the existence of
growing modes. Using (\ref{eq:defAB}) the latter inequality can be
equivalently expressed as $\xi>10/9$. Therefore, the intersection of
positive acceleration with the existence of growing modes finally
yields $\xi>2$. But this possibility is also unacceptable because
from Eq.\,(\ref{OmoOLoentropic1}) we realize that this would entail
$\Omo>2$ for entropic-force scenarios.
 As a result all entropic-force scenarios with values of $\Omo$ in a reasonable range
(in particular, the favorite range $\Omo\simeq 0.27-0.30$) are
incompatible with the existence of growing modes for structure
formation.

The upshot of this analysis is that there are no acceptable
phenomenological scenarios for the entropic-force cosmology in its
original formulation\,\cite{Easson10}. The model fails both at the
background and at the perturbations level. The last feature is
important for the following reason: if the failure at the background
level concerning the absence of a transition point from deceleration
to acceleration would be just the late-time effective behavior of a
more general model where the transition would occur, not even this
possibility would be allowed since the model has no growing modes
for the structure formation in the late-time universe. The core of
the problem stems once more from the structure of the DE evolution
law of the entropic-force model, Eq.(\ref{EffectiveDE}), which does
not have an additive constant and therefore has no $\CC$CDM limit
for any value of the parameter space.

\section{An alternative entropic model}
In the absence of a constant term in Eq.(\ref{EffectiveDE}) the
question that we want to address now is the following: {\it If we
replace the term $\CHd\,\Hd(t)$ with $C_{1}H(t)$ then is it possible
to provide a viable look-alike entropic-force model?} { Our aim is
to see if one can introduce a change in the structure of the model
(\ref{entropica}) such that at least one (or maybe the two) of the
previous problems can be fixed. Since higher powers of $H$ (say
$H^3$ or $H^4$ \cite{Perico}) can have no phenomenological
significance in the late universe (as they are too small), we used
just $H$. We were motivated also by the fact that vacuum models with
a linear term in $H$ have also been discussed previously in the
literature trying to explain the value of the cosmological constant
problem within QCD
\cite{SS2002,KlinkhamerVolovik09,Zhitnitsky09,Carn08}.}

In this case, the corresponding effective DE density is
\begin{equation}\label{EffectiveDE11}
\rDE(t)=\frac{3}{8\pi G}\left[C_{1}H(t)+\CH H^2(t)\,\right]\;,
\end{equation}
where here $C_1$ is of course a dimensionful new constant with the
same dimensions as $H$. Combining Eqs.(\ref{GeneralizedFriedmann}),
(\ref{Bronstein}) and (\ref{EffectiveDE11}) we find:
\begin{equation}\label{BasicdiffEquation1}
\dot{H}+\frac32\,\left(1+\wm\right)\left(1-C_{H}\right)H^2-\frac32\,C_{1}H
=0\,.
\end{equation}
The solution of it is \be H(t)=\frac{C_{1}}{\xim}\frac{{\rm
e}^{3C_{1}t/2}}{{\rm e}^{3C_{1}t/2}-1} \;\;, \label{frie4} \ee where
$\xim=\xi_{1}(1+\wm)$ with $\xi_{1}=1-\CH$. This parameter is
indicated distinctively with respect to the last section to make
clear that the structure of the model is indeed different and hence
the comparison with the data should produce in general different
numerical results.

Upon a new integration it is easy to prove that the scale factor of
the universe becomes \be a(t)=a_{1}\left({\rm e}^{3C_{1}t/2}-1
\right)^{2/3\xim} \;\;, \label{frie44} \ee where $a_{1}$ is the
constant of integration. Combining the above equations we are led to
a first expression for the Hubble function in terms of the scale
factor: \be
H(a)=\frac{C_{1}}{\xim}\left[1+\left(\frac{a}{a_{1}}\right)^{-3\xim/2}
\right] \;\;. \label{frie5} \ee

Below we focus on the matter dominated epoch namely, $\wm=0$ and
$\xim=\xi_{1}$. Evaluating Eq.(\ref{frie5}) at the present time
($a\equiv1$), we obtain \be C_{1}=\frac{\xi_{1}
H_{0}}{1+a_{1}^{3\xi_{1}/2}} \;\;. \label{frie6} \ee Since the
current value of the DE density (\ref{EffectiveDE11}) must match the
measured  value of the vacuum energy density
$\rDE(t_0)=\rLo=\OLo\rco=(1-\Omo) 3H_0^2/8\pi G$, we can determine
the coefficients
 \be \label{an1}
a_{1}=\left(\frac{\Omo}{\xi_{1}-\Omo}\right)^{2/3\xi_{1}}\,,
\;\;\;\; C_{1}=H_{0}(\xi_{1}-\Omo) \;\;. \ee
As we can see, the two natural independent parameters of this model
to be fitted to the data are $(\Omo,\xi_1)$.

Using the relations (\ref{an1}) in (\ref{frie5}) we obtain the
explicit form of the normalized Hubble parameter as a function of
the scale factor: \be
E(a)=\frac{H(a)}{H_{0}}=1-\frac{\Omo}{\xi_{1}}+
\frac{\Omo}{\xi_{1}}a^{-3\xi_{1}/2} \;\;. \label{frie7} \ee It
correctly satisfies $E(a=1)=1$ in the matter dominated
epoch\,\footnote{When we include later on the CMB shift parameter in
the statistical analysis we need to include the radiation component
in the Hubble function. We assume it strictly conserved, i.e.
$\rho_r=\rho_{r 0}\,a^{-4}$, with $\Omega_{r 0}=4.153\times
10^{-5}h^{-2}$ \cite{Shaf13} and $h=0.674$. We can effectively
include radiation in the context of Eq. (\ref{frie7}) by adding a
term $\Omega_{r0}\,(a^{-2}-1)$.}.
Obviously the following condition must hold: $\xi_{1}>\Omo$ (or
$C_{H}<\OLo$).

Finally, using the above equations the scale factor of the universe,
normalized to unity at the present epoch, becomes \be \label{frie8}
a(t)=\left(\frac{\Omo}{\xi_{1}-\Omo}\right)^{2/3\xi_{1}}\left[{\rm
e}^{3H_0\,(\xi_{1}-\Omo)\,t/2}-1 \right]^{2/3\xi_{1}} \;. \ee In the
model under consideration, there exist a transition epoch in which
the Hubble expansion changes from the decelerating to the
accelerating regime ($\ddot{a}=0$). This is in contrast to the model
considered in the previous section based on the DE density
(\ref{EffectiveDE}). For the current model the inflection point can
be readily calculated from the explicit form (\ref{frie7}) setting
$q=0$ in the  formula (\ref{eq:qnu}). In contrast to the previous
model, in this case it does not give a constant and hence it
determines the following value of the scale factor and corresponding
redshift:
\begin{eqnarray}
      \label{infle}
a_{I}=\frac{1}{1+z_{I}}=\left[\frac{(3\xi_{1}-2)\Omo}{2
(\xi_{1}-\Omo)}\right]^{2/3\xi_{1}}\,.
\end{eqnarray}
It becomes clear that the transition epoch is present only for
$\xi_{1} > 2/3$ (i.e. $C_H<1/3$)

While the model of the previous section was discarded both at the
background and perturbations level from pure analytical
considerations, in the present case the model cannot be ruled out on
the same grounds and we have to further check it by comparison with
the cosmological data.

Using a joint statistical analysis, involving the latest
observational data (SNIa-Union2.1 \cite{Suzuki:2011hu}, BAO
\cite{Blake:2011en,Perc10} and the Planck CMB shift parameter
\cite{Ade13,Shaf13}), an efficient test can be implemented. Notice
that the corresponding covariances can be found in Basilakos et al.
\cite{BasNes13} for the SNIa/BAO data and in \cite{Shaf13} for the
Planck CMB shift parameter respectively. We find that the overall
likelihood function peaks at $\Omo=0.296\pm 0.017$, $\xi_{1}=1.189
\pm 0.008$ with $\chi_{min}^{2}(\Omo,\xi_{1})\simeq 568.3$,
resulting in a reduced value of $\chi^{2}_{min}/dof\sim 0.97$. With
these numerical values we see from (\ref{an1}) that $C_1$ is of
order $H_0$, as could be expected.
At the same time we find that the transition epoch is taking place
at $a_{I} \sim 0.47$ which corresponds to a redshift $z_I\sim 1.13$.
This value is substantially higher than in the case of the
concordance $\Lambda$CDM model, namely $z_{I,\Lambda}\sim 0.70$.
However, this is not the main problem, we still have to attend the
analysis of matter perturbations.

\subsection{The growth factor}
In the light of the new growth data (as collected by
\cite{BasNes13}) we compare the growth of matter fluctuations of the
effective entropic-force model with observations. Following the
procedure of\, \cite{Bas09a,Bas09c}, to which we refer the reader
for more details, we introduce the new variable \be \label{tran1}
y={\rm exp}(3C_{1}t/2) \;\;{\rm with} \;\; 0<y<1 \ee and using
equations (\ref{GeneralizedFriedmann}), (\ref{defQ}) and
(\ref{EffectiveDE11}) we can write
\begin{eqnarray}\label{auxiliar1}
\rmm&=& \frac{3C^{2}_{1}y}{\xi_{1} (y-1)^{2}}\nonumber\\
{Q}&=&\frac{3C_{1}[(2-\xi_{1})y-\xi_{1}]}{2\xi_{1}(y-1)}\nonumber\\
\dot{Q}&=&\frac{9C^{2}_{1}(\xi_{1}-1)y}{2\xi_{1}(y-1)^{2}}\;\;.
\end{eqnarray}
Inserting Eqs.(\ref{tran1}) and (\ref{auxiliar1}) into
Eq.(\ref{diffeqD}) we arrive at the following differential equation:
\begin{eqnarray}
\label{eq:22} 3\xi_{1}^{2} y(y-1)^{2}D^{''}+2\xi_{1}
(y-1)(5y-3\xi_{1})D^{'}\nonumber\\-2(2-\xi_{1})(3\xi_{1}-2y)D=0\,,
\end{eqnarray}
where primes indicate derivatives with respect to the new variable
$y$. The latter is related with the scale factor as follows: \be
y=1+\frac{\xi_{1}-\Omo}{\Omo}\,a^{3\xi_{1}/2} \,.
\ee
The decaying solution of Eq.(\ref{eq:22}) can be identified easily,
it reads $D_{-}(y)=(y-1)^{(\xi_{1}-2)/\xi_{1}}\sim
a^{3(\xi_{1}-2)/2}$  for $\xi_1<8/3$ (hence $C_H>-5/3$). Thus, the
corresponding growing mode of Eq.(\ref{eq:22}) is \be \label{eqff2}
D_{+}(a)=C a^{3(\xi_{1}-2)/2}\int_{0}^{a} \frac{{\rm
d}x}{x^{3\xi_{1}/2}E^{2}(x)} \ee where \be
C=\frac{3\Omo^2}{2\xi_1}\left(\frac{\xi_1-\Omo}{\Omo}\right)^{2(3\xi_{1}-2)/3\xi_{1}}
\ee

\begin{figure}
\mbox{\epsfxsize=14.2cm \epsffile{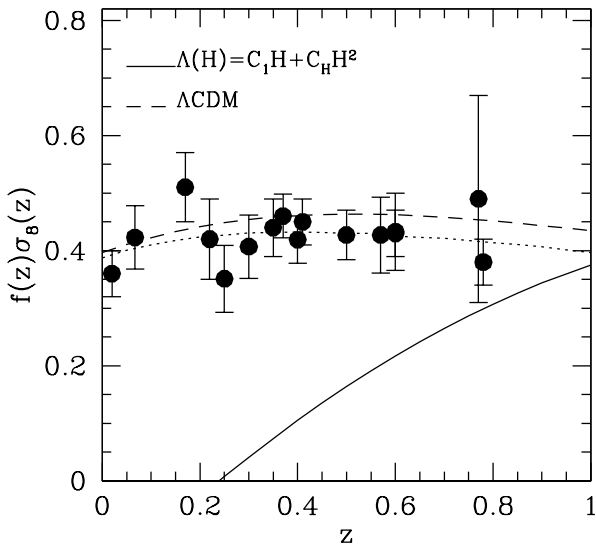}} \caption{Comparison of
the observed (solid points) and theoretical evolution of the growth
rate $f(z)\sigma_{8}(z)$. The solid line corresponds to the
alternate entropic-force DE model (\ref{EffectiveDE11}) for the best
fit values $(\Omega_{m0},\xi_{1})=(0.296,1.189)$ discussed in Sec.4.
The dotted line shows the predicted growth rate in the case of
$(\Omega_{m0},\xi_{1})=(0.73,1.01)$. For comparison we also plot the
$\Lambda$CDM (dashed line). { We use $\sigma_{8}=0.818$
\cite{Sperg2013} while for the $\Lambda$CDM case we set}
$\Omo=0.272$ \cite{BasNes13}.}
\end{figure}

We point out that in the limit when the linear term in the Hubble
function of the DE density (\ref{EffectiveDE11}) vanishes ($C_1\to
0$) the above formulas simplify dramatically. In this case the
cosmological solution takes on the much simpler form:
\begin{equation}\label{C10}
E(a)\sim a^{-3\xi_{1}/2}\,,\ \ \ \  D_{+}(a)\sim a^{3\xi_{1}-2}.
\end{equation}
The resulting DE model is of course coincident with the situation
$\CHd=0$ studied for the model of Section 3, which was problematic
because the existence of growing modes is incompatible with having
accelerated expansion, see Eq.\,(\ref{eq:solperturbH2}).

Finally, let us consider the evolution of the growth rate of
clustering, defined as $f(a)=d{\rm ln}D_{+}/d{\rm ln}a$. It can be
computed from the growing mode solution (\ref{eqff2}). We find: \be
f(a)=\frac{3(\xi_{1}-2)}{2}+\frac{C}{a^{2}E^{2}(a)D_{+}(a)} \;. \ee
In Fig.1 we plot the growth data as collected by Basilakos et al.
(see \cite{BasNes13} and references therein) with the estimated
growth rate function, $f(z)\sigma_{8}(z)$. The  solid line in the
figure corresponds to the aforementioned best fit values for
SNIa/BAO data and the Planck CMB shift parameter, i.e. for
$(\Omega_{m0},\xi_{1})=(0.296,1.188)$.
 We have included also the prediction of the concordance $\Lambda$CDM model -- see the dashed line.
Note, that the theoretical $\sigma_{8}(z)$ is given by
$\sigma_{8}(z)=\sigma_{8}D(z)$, where $D(z)=D_{+}(z)/D_{+}(z=0)$ is
the growth factor scaled to unity at the present time, and
$\sigma_{8}$ is the rms mass fluctuation on $R_{8}=8 h^{-1}$ Mpc
scales at redshift $z=0$. {We use for it the Planck parametrization
value: $\sigma_{8}=0.818$ \cite{Sperg2013}.}

From the comparison it becomes clear that the present growth data
strongly disfavors the alternate entropic-force DE model
(\ref{EffectiveDE11}). The lack of structure formation near our time
is quite evident as compared to the concordance $\CC$CDM model. If
we, however, enforce that the model optimally fits only the data on
the growth rate we find: $\xi_{1}\simeq 1.01$ and $\Omega_{m0}\simeq
0.73$ (see dotted line in Fig.1).
Obviously, the obtained value of $\Omega_{m0}$ is far beyond the
acceptable physical range of values and thus it must be rejected.

Let us also mention that the DE model (\ref{EffectiveDE11}) with
$\xi_{1} \to 1$ (i.e. $C_{H}\to 0$) reduces to that of Borges et al.
\cite{Borg, Carn08} in which the DE term is directly proportional to
the Hubble parameter, $\rDE(a) \propto H(a)$. In this framework, we
repeat our joint statistical analysis by imposing $\xi_{1}=1$ into
Eq.(\ref{frie7}) and we find that the overall fit provides
$\Omega_{m0}=0.30 \pm 0.01$, but with a poor quality:
$\chi_{min}^{2}(\Omega_{m0})/dof\simeq 2.7$.
The main problem here is related with the fact that the $\rDE(a)
\propto H(a)$ model can not accommodate the CMB data. Indeed, using
the Planck CMB shift parameter alone we find that the corresponding
likelihood function peaks at the totally unrealistic value
$\Omega_{m0}\simeq 0.97$, which is $\sim 3.2$ times larger than that
provided by the SNIa+BAO solution $\Omega_{m0} \simeq 0.30$.

Finally, the following observation is in order. The DE models in
which the energy density is strongly dependent on odd powers of the
Hubble rate are not favored from the theoretical point of
view\,\cite{RGCosmology1,RGCosmology2,Fossil07} as it is impossible
to fit them into the covariant form of the effective action of QFT
in curved spacetime. If they contain a combination of even and odd
powers of $H$, one can treat the odd powers phenomenologically as
representing e.g. bulk viscosity effects. However only the even
powers can have a fundamental origin. In this sense DE models based
on pure linear terms in $H$ are not well motivated, and in fact we
find they are poorly fitting the main cosmological data. We have
seen that after including the $H^2$ term in it, as in
Eq.\,(\ref{EffectiveDE}), the new DE density behaves better at the
background level since it provides an acceptable fit value around
$\Omo\sim 0.3$. Notwithstanding, it does not pass the stringent test
of the structure formation data.

Overall, the above analysis points that even if we replace the term
$\CHd\,\Hd(t)$ in Eq.(\ref{EffectiveDE}) with $C_{1}H(t)$ the
modified DE model, namely Eq.\,(\ref{EffectiveDE11}), is unable to
fit simultaneously the observational data on Hubble expansion and
structure formation.

\section{Conclusions}
In this work we have considered the  entropic-force
model\,\cite{Verlinde10} and reanalyzed the implications as a
possible dark energy candidate in cosmology within the context of
\,\cite{Easson10}. We have found that in its original formulation
the effective dark energy implied by such model is not viable
neither at the background nor at the cosmic perturbations level.
Specifically, the analysis of the deceleration parameter immediately
detects a fundamental problem, to wit: the expansion of the universe
always accelerates or always decelerates. While this is of course a
severe problem it might still be cured if such behavior were only
the late-time behavior of a more complete model where the transition
from deceleration to acceleration would be present. However, a
detailed analysis of the cosmic perturbations furnishes also a
negative result, and this may give the final blow to the model: it
turns out that the existence of growing modes for structure
formation is incompatible with the accelerated expansion.

In an attempt to rescue some form of the entropic-force DE model, we
have also addressed a modification of its energy density by
replacing the time derivative term $\dot{H}$ with the linear term
$H$ while keeping the $H^2$ one. The modified model is better
behaved at the background level, for there is a transition point
from deceleration to acceleration and moreover the best fit value of
$\Omo$ is near $0.3$. However, the modified model is once more
unable to fit the most recent growth data on structure formation.
Let us mention that if we would also suppress the $H^2$ component
(i.e. if the DE density would be reduced to the linear term in the
Hubble function), then the best fit value of $\Omo$ has poor quality
in the allowed region and runs completely out of range (near $1$) if
we single out the CMB data.

In all these cases the absence of an additive term in the structure
of the DE density lies at the root of the main difficulties with
these models. This constant is essential for the transition from
deceleration into acceleration. It is also essential to provide a
late-time linear growth rate that can describe the structure
formation data in a way comparable to the $\CC$CDM model. If the
structure of the entropic-force models is corrected with an additive
term, i.e. if the DE density becomes an ``affine function'' of the
Hubble terms, $\rDE=c_0+c_1\,H+c_{\dot{H}}\,\dot{H}+c_2\,H^2$, then
these models have a well-defined $\CC$CDM limit when the
coefficients of the $H$,$\dot{H}$ and $H^2$ terms approach zero.
These modified models can be phenomenologically compatible with the
cosmological data\,\cite{Polarski,Adria}.

Let us note that we have focused on the study of the entropic-force
models of the $\CC$-type, i.e. such that the DE density
(\ref{EffectiveDE}) has the equation of state $\wDE=-1$, as in
Ref.\cite{Polarski}. One may consider other possible
implementations\,\cite{Japanese12,Japanese3}, but to get a viable
one it is unavoidable to introduce in all cases an additive term.
Unfortunately this cannot be done in a natural way in the
entropic-force formulation.

Such situation is in stark contradistinction to the running vacuum
models based on the renormalization group in semiclassical gravity
in curved spacetime\,\cite{JSP-CCReview13}. The latter types of
models exist in the literature since long ago -- see
\,\cite{RGCosmology1,RGCosmology2,Fossil07} and references therein
-- and have been successfully
tested\,\cite{Bas09c,GSBP11,PerturbVacuum1,PerturbVacuum2}. They
naturally incorporate the additive term as an integration constant,
namely in the ``affine'' form mentioned above, but only the even
powers of $H$ (and the powers of $\dot{H}$) are allowed, as expected
from the covariant QFT formulation in a curved background.

The entropic-force models that became popular lately are actually a
particular case of the running vacuum models, as clearly discussed
in \,\cite{Polarski}. This particular case, however, proves to be
phenomenologically inviable at the background and/or at the
perturbations level, as we have clearly demonstrated here. Even if
other formulations of the entropic models are possible the main
result, namely that simple combinations of pure Hubble terms
$H$,$\dot{H}$, $H^2$ are not sufficient for a complete description
of the cosmological data, stays in force.

\section*{Acknowledgments}
We thank T. Padmanabhan for useful comments. SB acknowledges support
by the Research Center for Astronomy of the Academy of Athens in the
context of the program {\it ``Tracing the Cosmic Acceleration''}. JS
has been supported in part by FPA2010-20807 (MICINN), Consolider
grant CSD2007-00042 (CPAN) and by DIUE/CUR Generalitat de Catalunya
under project 2009SGR502.


\begin{thebibliography}{99}

\bibitem {Spergel07}
D.N. Spergel, et al., Astrophys. J. Suplem., \textbf{170}, 377,
(2007)

\bibitem {essence}
T.M. Davis \textit{et al.}, Astrophys. J., \textbf{666}, 716, (2007)

\bibitem {Kowal08}
M. Kowalski, et al., Astrophys. J., \textbf{686}, 749, (2008)

\bibitem {Hic09}
M. Hicken et al., Astroplys. J., \textbf{700}, 1097, (2009)

\bibitem {komatsu08}
E. Komatsu, et al., Astrophys. J. Suplem., \textbf{180}, 330,
(2009); G. Hinshaw, et al., Astrophys. J. Suplem., \textbf{180},
225, (2009); E. Komatsu, et al., Astrophys. J. Suplem.,
\textbf{192}, 18, (2011)

\bibitem {LJC09}
J. A. S. Lima and J. S. Alcaniz, Mon. Not. Roy. Astron. Soc.
\textbf{317}, 893 (2000); J. F. Jesus and J. V. Cunha, Astrophys. J.
Lett. \textbf{690}, L85 (2009)

\bibitem {BasPli10}S. Basilakos and M. Plionis, Astrophys. J. Lett,
\textbf{714}, 185 (2010)

\bibitem{Ade13}
P. A. R. Ade, et al. (Planck Collaboration), (2013),
  [arXiv:1303.5076]

\bibitem{Sperg2013}
D. Spergel, R. Flauger and R. Hlozek, (2013), [arXiv:1312.3313]

\bibitem{CCproblem} S. Weinberg, Rev. Mod. Phys. {\bf 61} (1989) 1; V. Sahni, A. Starobinsky, Int. J. of Mod. Phys. {\bf
 A9} {(2000)} {373};
T. Padmanabhan, Phys. Rept.  {\bf 380} (2003)  235.

\bibitem{JSP-CCReview13} J. Sol\`a,  \textit{Cosmological constant and vacuum energy: old and new ideas}, J. Phys. Conf. Ser. (2013) {\bf 453} 012015 [arXiv:1306.1527]; \textit{
Vacuum energy and cosmological evolution},  e-Print:
arXiv:1402.7049.

    \bibitem{Dolgov82} A.D.~Dolgov, in: \textit{The very Early Universe}, Ed.
    G.~Gibbons, S.W.~Hawking, S.T.~Tiklos (Cambridge U., 1982); Y. Fujii, Phys. Rev. {\bf D26} (1982) 2580.

 \bibitem{OldDynamAdjust}    L. Abbott, {Phys. Lett.} \textbf{B150} (1985) 427; S. M. Barr, Phys. Rev. D {\bf 36}{1987}{1691}; L.H. Ford, Phys.
    Rev. {\bf D35} (1987) 2339; E.T. Tomboulis, Nucl.Phys. {\bf B329}
(1990) 410.

\bibitem{PSW} R.D.~Peccei, J.~Sol\`{a} and C.~Wetterich,
Phys. Lett. {\bf B195} {(1987)} {183}; J.~Sol\`{a}, Phys. Lett. B228
(1989) 317;  Int.J.  Mod. Phys.  {\bf A5} (1990) 4225.

\bibitem{Quintessence}  C. Wetterich, Nucl. Phys. B {\bf 302} {1988} 668;
B. Ratra, P.J.E. Peebles, Phys. Rev. D {\bf 37} {1988}
    {3406};
C. Wetterich, Astron. Astrophys. \textbf{301}, 321 (1995);
R.R. Caldwell, R. Dave, P.J. Steinhardt, Phys. Rev. Lett. {\bf 80}
(1998) 1582;


\bibitem{QuintessenceReview} P.J. Peebles, B. Ratra, Rev. Mod. Phys. {\bf 75} (2003) 559.


\bibitem {curvature}
S. Capozziello, Int.J.Mod.Phys. \textbf{D 11}, 483 (2002).

\bibitem {mauro}
S. Capozziello, M. Francaviglia Gen.Rel.Grav. \textbf{40}, 357
(2008).

\bibitem {report}
S. Capozziello, M. De Laurentis, Phys. Rept. \textbf{509}, 167
(2011).

\bibitem{repsergei}
S. Nojiri, S. D. Odintsov , Phys.Rept. \textbf{505}, 59 (2011).

\bibitem{Oze87}
M. Ozer and O. Taha, Nucl. Phys. B \textbf{287}, 776 (1987).

\bibitem {Lambdat}W. Chen and Y-S. Wu, Phys. Rev. D \textbf{41}, 695 (1990);
J. C. Carvalho, J. A. S. Lima and I. Waga, Phys. Rev. D
\textbf{{46}}, 2404 (1992); J. A. S. Lima and J. M. F. Maia, Phys.
Rev D \textbf{49}, 5597 (1994); J. A. S. Lima, Phys. Rev. D
\textbf{54}, 2571 (1996); A. I. Arbab and A. M. M. Abdel-Rahman,
Phys. Rev. D \textbf{50}, 7725 (1994); J. M. Overduin and F. I.
Cooperstock, Phys. Rev. D \textbf{{58}}, 043506 (1998).

\bibitem{Brax:1999gp}
P. Brax, and J. Martin, Phys. Lett. \textbf{B468}, 40 (1999)

\bibitem{KAM}
A. Kamenshchik, U. Moschella, and V. Pasquier, Phys. Lett. B.
\textbf{511}, 265, (2001)

\bibitem {fein02}
A. Feinstein, Phys. Rev. D., \textbf{66}, 063511 (2002)

\bibitem {Caldwell}
R. R. Caldwell, Phys. Rev. Lett. B., \textbf{545}, 23 (2002)

\bibitem {Bento03}
M. C. Bento, O. Bertolami, and A.A. Sen, Phys. Rev. D., \textbf{70},
083519 (2004)

\bibitem {chime04}
L. P. Chimento, and A. Feinstein, Mod. Phys. Lett. A, \textbf{19},
761 (2004)

\bibitem {Linder2004}
E. V. Linder, Rep.\ Prog.\ Phys., \textbf{71}, 056901 (2008)

\bibitem {LSS08}
J. A. S. Lima, F. E. Silva and R. C. Santos, Class. Quant. Grav.
\textbf{25}, 205006 (2008)

\bibitem {Brookfield:2005td}
A. W. Brookfield, C. van de Bruck, D.F. Mota, and D.
Tocchini-Valentini, Phys. Rev. Lett. \textbf{96}, 061301 (2006)

\bibitem {Boehmer:2007qa}
C.~G. Boehmer, and T.~Harko, Eur. Phys. J. \textbf{C50}, 423 (2007)

\bibitem {Starobinsky-2007}
A.~A.~Starobinsky, JETP Lett.\ \textbf{86}, 157 (2007)

\bibitem{Ame10}E. J. Copeland, M. Sami and S. Tsujikawa, Intern. J. of
Modern Physics D, \textbf{15}, 1753,(2006); R. R. Caldwell and M.
Kamionkowski, Ann.Rev.Nucl.Part.Sci., \textbf{59}, 397, (2009),
arXiv:0903.0866; I. Sawicki and W. Hu, Phys. Rev. D., \textbf{75},
127502 (2007); L. Amendola and S. Tsujikawa, Dark Energy Theory and
Observations, Cambridge University Press, Cambridge UK, (2010); S.
Capozziello and V. Faraoni, Beyond Einstein gravity: A Survey of
gravitational theories for cosmology and astrophysics, Fundamental
Theories of Physics, Vol. 170, Springer, Heidelberg (2010).

\bibitem{RGCosmology1} I.~L. Shapiro and J.~Sol{\`a}, JHEP {\bf 02}, 006
    (2002) [arXiv:hep-th/0012227];  Phys. Lett. B
    {\bf 475}, 236 (2000) [arXiv:hep-ph/9910462]; {Nucl. Phys. Proc.
    Suppl.} {\bf 127} {(2004)} {71} [hep-ph/0305279]; JHEP proc. AHEP2003/013 [astro-ph/0401015].

\bibitem{RGCosmology2}  I. L. Shapiro and J. Sol{\`a},  Phys. Lett. B {\bf 682},  105 (2009) [arXiv:0910.4925]. See also the extended discussion in I. L. Shapiro and J. Sol{\`a}, \textit{Can the cosmological 'constant' run? - It may run}, e-Print: [arXiv:0808.0315].

\bibitem{Fossil07} J.~Sol{\`a}, J. Phys. A {\bf 41} (2008) 164066 [arXiv:0710.4151].

\bibitem{Bas09c}
S. Basilakos, M. Plionis and J. Sol\`{a}, Phys. Rev. D. \textbf{80},
3511 (2009) [arXiv:0907.4555].

\bibitem{GSBP11}
J. Grande, J. Sol\`{a}, S. Basilakos and M. Plionis, JCAP {\bf 08},
007 (2011) [arXiv:1103.4632];


\bibitem{Verlinde10}
E.P. Verlinde, JHEP {\bf 04}, 029, 2011 [arXiv:1001.0785].


\bibitem{Easson10}
D. A. Easson, P. H. Frampton, G.F. Smoot,  \PLB {696}{273}{(2011)}
[arXiv:1002.4278].

\bibitem{Easson10b}
D. A. Easson, P. H. Frampton, G.F. Smoot, \textit{Entropic
Inflation}, Int.J.Mod.Phys. A {\bf 27} (2012) 1250066
[arXiv:1003.1528].

\bibitem{Jacobson95}
T. Jacobson, Phys. Rev. Lett., {\bf 75}, 1260 (1995)


\bibitem{Padmanabhan10} T. Padmanabhan,
Rept. Prog. Phys. {\bf 73} (2010) 046901 [arXiv:0911.5004]; \PR
{406}{49}{(2005)}

\bibitem{Visser11} M. Visser, \JHEP {10} {140} {(2011)}  [arXiv:1108.5240].

\bibitem{Mia011}
Miao Li and Yi Wang, Phys. Lett. B., {\bf 687}, 243 (2010)

\bibitem{Casadio10}
R. Casadio, A. Gruppuso, Phys. Rev. D. {\bf 84}, 023503, (2011); T.
S. Koivisto, D. F. Mota, M. Zumalacarregui, J. of Cosmology and
Astrop. Phys., {\bf 02}, 027 (2011).


\bibitem{Entropic-others}
Y-F. Cai, J. Liu, H. Li, \PLB {690}{213}{(2010)};
H. Wei \PLB {692}{167} {(2010)};\\
Y.S.  Myung, Astrophys. Space Sci. {\bf 335} (2011) 553;
Y. Fu Cai, E. N. Saridakis, \PLB {697} {280} {(2011)}


\bibitem{Polarski}
S. Basilakos, D. Polarski and J. Sol\`a,  Phys. Rev. D{\bf 86},
043010 (2012) [arXiv:1204.4806].


\bibitem{Japanese12}
N. Komatsu, S. Kimura [arXiv:1402.3755]
Phys.Rev. {\bf D88} (2013) 083534 [arXiv:1307.5949];
Phys.Rev. {\bf D87} (2013) 043531 [arXiv:1208.2482].

\bibitem{Japanese3}
N. Komatsu, S. Kimura [arXiv:1402.3755].

\bibitem{Hawking96}
S.W. Hawking, G. T. Horowitz, \CQG {13} {1487} {(1996)}

\bibitem{Grande08} J. Grande, A. Pelinson, J. Sol\`{a},
Phys. Rev. D. {\bf 79}, {043006}, {(2009)} [arXiv:0809.3462].

\bibitem{Arc94}
R. C. Arcuri and I. Waga., Phys. Rev. D., {\bf 50}, 2928, (1994)

\bibitem{Borg}
H. A. Borges, S. Carneiro, J. C. Fabris and  C. Pigozzo, Phys. Rev.
D., {\bf 77}, 043513 (2008).


\bibitem{PerturbVacuum1} J. Grande, J. Sol\`a, J.C. Fabris, I.L.Shapiro, {Class. Quant. Grav.} {\bf 27} (2010) 105004 [arXiv:1001.0259];

\bibitem{PerturbVacuum2} J. C. Fabris, I. L. Shapiro, J. Sol\`a, JCAP 0702 (2007) 016 [arXiv:gr-qc/0609017]; A. M. Velasquez-Toribio,  Int. J. of Mod Phys. {\bf D21} (2012) 1250026 [arXiv:0907.3518].

\bibitem{PeeblesDodelson} P. J. E. Peebles, \textit{Principles of Physical Cosmology} (Princeton
Univ. Press, Princeton New Jersey, 1993); S. Dodelson,
\textit{Modern Cosmology} (Academic Press, 2003).


\bibitem{Perico}
J.~A.~S. Lima, S.~Basilakos, and J.~Sol{\`a},
    Mon. Not. Roy. Astron. Soc. {\bf 431} (2013) 923 [arXiv:1209.2802]; E. L. D. Perico, J.A.S. Lima, S.~Basilakos, and J.~Sol{\`a},
Phys. Rev. {\bf D88} (2013) 063531 [arXiv:1306.0591];
    S.~Basilakos,  J.~A.~S. Lima, and J.~Sol{\`a},  Int. J. Mod. Phys. D (2013) [arXiv:1307.6251].

\bibitem{SS2002}
R. Schutzhold, Phys. Rev. D., {\bf 89}, 081302 (2002)


\bibitem{KlinkhamerVolovik09}
F.R. Klinkhamer, G.E. Volovik, Phys.Rev. {\bf D79} (2009) 063527.

\bibitem{Zhitnitsky09} E. C. Thomas, F. R. Urban, A. R. Zhitnitsky, JHEP {\bf 0908} (2009) 043;
 N. Ohta, Phys. Lett. {\bf B 695} (2011) 41.

\bibitem{Carn08}
S. Carneiro, M. A. Dantas, C. Pigozzo and J. S. Alcaniz, Phys. Rev.
D., {\bf 77}, 083504, (2008).



\bibitem{Suzuki:2011hu}
  N.~Suzuki, D.~Rubin, C.~Lidman, G.~Aldering, R.~Amanullah,  K.~Barbary,
L.~F.~Barrientos and J.~Botyanszki {\it et al.},
  Astrophys.\ J\  {\bf 746}, 85 (2012).

\bibitem{Blake:2011en}
  C.~Blake, E.~Kazin, F.~Beutler, T.~Davis, D.~Parkinson, S.~Brough,
M.~Colless and C.~Contreras {\it et al.},
  Mon.\ Not.\ Roy.\ Astron.\ Soc.\  {\bf 418}, 1707 (2011).

\bibitem{Perc10} W. J. Percival, Mon. Not. Roy. Astron. Soc., {\bf 401}
2148 (2010)


\bibitem{Shaf13}
D. L. Shaefer and D. Huterer, (2013), [arXiv:1312.1688]

\bibitem{BasNes13}
  S.~Basilakos, S.~Nesseris and L.~Perivolaropoulos,
Phys.\ Rev.\ D {\bf 87}, 123529 (2013).

\bibitem{Bas09a}
S. Basilakos, Mon. Not. Roy. Astron. Soc., {\bf 395}, 2347, (2009a)

\bibitem{Adria}
S. Basilakos, A. G\'omez, and J. Sol\`a, in preparation.


\end{thebibliography}
\end{document}